\documentclass[12pt,preprint]{aastex}
\usepackage{psfig}
\def\gsim{\lower 2pt \hbox{$\, \buildrel {\scriptstyle >}\over
{\scriptstyle \sim}\,$}}
\def\lsim{\lower 2pt \hbox{$\, \buildrel {\scriptstyle <}\over
{\scriptstyle \sim}\,$}}

\def\rosat{{\sl ROSAT}}
\def\asca{{\sl ASCA}}
\def\chandra{{\sl Chandra}}
\def\Chandra{{\sl Chandra}}

\def\snr{{N157B}}
\def\psr{{PSR J0537-6910}}

\shortauthors{Wang, Gotthelf, Chu, \& Dickel}
\shorttitle{X-ray nebula around \psr}

\begin{document}

\title{Detection of an X-ray Pulsar Wind Nebula and Tail in SNR N157B}

\author{Q. D. Wang}
\affil{Department of Astronomy, University of Massachusetts}
\affil{524 LGRT, Amherst, MA 01003}
\affil{Electronic mail: wqd@astro.umass.edu}

\author{E. V. Gotthelf}
\affil{Columbia Astrophysics Laboratory, Columbia University}
\affil{550 West 120 St, New York,~NY, 10027}
\affil{Electronic mail: evg@astro.columbia.edu}
\author{Y.-H. Chu and J. R. Dickel}
\affil{Astronomy Department, University of Illinois}
\affil{1002 West Green Street, Urbana, IL 61801}
\affil{Electronic mail: chu@astro.uiuc.edu and johnd@astro.uiuc.edu}

\begin{abstract}

We report {\sl Chandra} X-ray observations of the supernova remnant
\snr~in the Large Magellanic Cloud, which are presented together with
an archival {\sl HST} optical image and a radio continuum map for
comparison. This remnant contains the recently discovered 16 ms X-ray pulsar
  PSR~J0537-6910, the most rapidly rotating young pulsar known. 
  Using phase-resolved Chandra imaging, we pinpoint the location of the
  pulsar to $5^h37^m47\fs36, -69^\circ 10^\prime$20\farcs4 (J2000)
  with an uncertainty of $\lsim 1^{\prime\prime}$. PSR~J0537-6910 is not
  detected in any other wavelength band. The X-ray observations
resolve three distinct features: the pulsar itself, a surrounding
compact wind nebula which is strongly elongated (dimension $\sim 0.6$
pc $\times 1.7$ pc), and a feature of large-scale ($\gsim 5$ pc long)
diffuse emission trailing from the pulsar.  This
latter comet tail-shaped feature coexists with enhanced radio emission and
is oriented nearly perpendicular to the major axis of the pulsar wind
nebula.

We propose the following scenario to explain these features. The
bright, compact nebula is likely powered by a toroidal pulsar wind of
relativistic particles which is partially confined by the ram-pressure
from the supersonic motion of the pulsar. The particles, after being
forced out from the compact nebula (the head of the ``comet''), are
eventually dumped into a bubble (the tail), which is primarily responsible for
the extended diffuse X-ray and radio emission.  The ram-pressure
confinement also allows a natural explanation for the observed X-ray
luminosity of the compact nebula and for the unusually small X-ray to
spin-down luminosity ratio of $\sim 0.2\%$, compared to similarly
energetic pulsars.  We estimate the pulsar wind Lorentz factor of
N157B as $\sim 4 \times 10^6$ (with an uncertainty of a factor $\sim 2$), 
consistent with that inferred from the modeling of the Crab Nebula.

\end{abstract}


\section{Introduction}

	Pulsars are thought to lose their rotation energy
predominately in the form of a highly relativistic electron/positron
wind.  Initially, the wind is invisible as it freely expands through
the self evacuated region surrounding the pulsar, and eventually
encounters ambient medium where the wind is reverse-shocked, resulting in
the thermalization and re-acceleration of particles. The synchrotron
radiation from these particles may be strong enough to be observable.
The classic example of such a pulsar wind nebulae (PWNe) is the Crab
nebula (containing a pulsar with a spin period of 33 ms) and the supernova 
remnant (SNR) 0540-693 in the Large Magellenic Cloud (LMC) which
contains a 50 ms pulsar (Gotthelf \& Wang 2000 and references therein).
Extracting information from a Crab-like PWN is difficult though,
because their present physical condition depends on their evolutionary
history, which can be rather uncertain.

	Optimally, one wants to study a PWN that is strongly confined by ram
pressure.  Such confinement occurs when the pulsar
responsible for the PWN is moving at a velocity higher than the sound
speed of the ambient medium. In this case, the shocked wind assumes a
bow-shock morphology. Driven by the high pressure inside the bow shock
region, shocked wind material can flow out in the direction {\it
opposite} to the pulsar's proper motion (e.g., Wang et al. 1993; Wang
\& Gotthelf 1998a --- WG98 hereafter). Consequently, the bow shock
region contains only freshly shocked wind material. Thus the
synchrotron radiation (primarily in X-ray) from this region contains
the most direct information about both the pre-shock pulsar wind and
the subsequent particle acceleration in the shock. The moderate
relativistic outflow will eventually be terminated by a weak reverse
shock or a series of oblique shocks, forming a
pulsar wind bubble that is largely offset from the pulsar (Wang et al. 1993).

The bright X-ray object N157B (NGC 2060, Henize 1956; also known as
SNR 0538-691) is a young (about $5,000$ yr-old; WG98) supernova remnant in
the 30 Doradus region of the Large Magellanic Cloud (LMC) with a
complex X-ray morphology. To explain the nature of the emission from
this object, WG98 proposed the ram-pressure confined PWN model, based
on a spatial and spectral analysis of \rosat\ and \asca\
observations. The subsequent discovery of the 16 ms pulsar \psr\ in
the remnant (Marshall et al. 1998) validated this interpretation and a
later \rosat\ HRI analysis located the pulsar emission to where
predicted by the model (Wang \& Gotthelf 1998b). The \rosat\ data also
marginally resolved large-scale diffuse X-ray emission extending out
from one side of the pulsar. The data, however, did not allow for a
spatial separation of the expected bow shock nebula from the pulsar.

The \chandra\ observations reported here enabled us to conduct a high
spatial resolution X-ray study of N157B.  We further compare the X-ray
data with an archival {\sl HST} WFPC2 optical image and
a radio continuum map.  Our results are consistent with the confined
PWN interpretation and further allow us to infer the geometry and
Lorentz factor of the pulsar wind. Throughout this paper, we adopt a
distance of N157B as 51 kpc (thus $1^{\prime\prime}$ corresponds to
0.25 pc).

\section {Observations and Calibration}

N157B was imaged by the
\Chandra\ observatory (Weisskopf et al. 1996) twice, once with each of its
two High Resolution Cameras, the HRC-I and HRC-S (Murray et al. 1997),
placed at the focal plane of the telescope mirror. The original HRC-I
observation, taken on 12 Feb 2000, suffered from a non-recoverable timing
problem (see the Chandra Science Center memo for details) which 
prevented the detection of the 
expected pulse period from \psr, whose pulse profile is just 2 ms wide. 
A re-observation, obtained on 30 July 2000 with the HRC-S, was carried
 out in a revised mode designed to bypass the timing problem of the HRCs.
  This observation served as the
verification of this new observing mode which is now standard for HRC
targets that require pulse timing uncertainty of less than 4 ms.  

The HRCs are multi-channel plate detectors sensitive to X-rays over
an energy range of $0.1-10.0$ keV with essentially no spectral information
available. For both observations, N157B was centered on the on-axis position 
of the
HRCs where the point spread function (PSF) has a half power radius (the
radius enclosing 50\% of total source counts) of $\sim 0\farcs5$.
The detectors oversample the telescope mirror PSF
 with pixels 0\farcs1318 per side. Only the HRC-S observation was used in 
the timing analysis reported herein. Time-tagged photons were acquired with 
a nominal precision of 15.6 $\mu$s,
and the arrival times were corrected to the solar system barycenter using
a beta version of {\tt AXBARY}. The initial pulsar position for the
barycenter correction was estimated using the reported \rosat\ position
(Wang \& Gotthelf 1998b).

Having previously identified various problems with the standard processed
HRC data sets, we begin our analysis with photon event data re-processed
from the Level 0.5 event files according to the prescription and filtering
criteria outlined in Helfand, Gotthelf, \& Halpern (2001). This reduced
the instrumental background, removed ``ghost image'' artifacts, applied
the correct degap parameters, filtered out time intervals of telemetry
dropouts, and reassigned mis-placed photons. The filtered/total exposure
times were 26.6/30.4 ks and 26.4/32.5 ks, respectively, for the first and
second observations. 

Both data sets are useful. The HRC-I field of view covered the core
region of 30 Dor, which harbors two relatively bright X-ray sources
that are apparently associated with Wolf-Rayet stars, MK34 and R140
(Wang 1995). The HRC-I centroids of the two X-ray sources are
found to be consistent with the optical positions of these stars; the offset is
$\lsim 1^{\prime\prime}$. Thus, we consider the astrometry of the
HRC-I observation to be accurate. However, the HRC-S centroid of \psr\
is offset by $\sim 2\farcs5$ from the measured HRC-I centroid of the
pulsar. Since the fiducial stars were out of the HRC-S field of view
and thus unavailable, we aspected the HRC-S observation using the
pulsar's centroid to the HRC-I coordinates. The two observations were
then combined to increase the counting statistics for images used to
produce Figs. 1-3. Since the reponses of the two instruments are not 
necessarily identical, all the our quantitative measurements reported
here are based on the HRC-S data alone.

\begin{figure} 
\centerline{ {\hfil\hfil
\hfil\hfil}}
\caption{{\sl Chandra} X-ray image of SNR N157B. The image,
a combination of the HRC-I and -S observations, is smoothed with 
an adaptive filter (program CSMOOTH implemented in the CIAO software package) 
and the chosen signal-to-noise ratio is 2.5. The contours
are at 0.85, 1.0, 1.1, 1.2, 1.4, 1.8, 2.2, 3, 4, 6, 8, 11, 15, 20, 26, 40, 
80, 160, 320, and 700 $\times 10^{-5} {\rm~counts~s^{-1}~arcsec^{-2}}$.
\label{fig1}}
\end{figure}

For a comparison with the \chandra\ data, we downloaded an archival 
{\sl HST} WFPC2 image, which was taken on 1997 October 16 (Proposal ID 7786).  
As this {\sl HST} observation was a single 40-sec exposure, the 
cosmic-ray-removal program was not performed on the image (Fig.\ 2).  
The apparent linear artifacts were caused by cosmic ray hits.  
The correction of geometric distortion was not
applied on the image, which can result in an error of about 0\farcs1 on 
the relative positions.  The accuracy of absolute astrometry on WFPC2
images, limited by the accuracy of the guide star position, is 
typically 0\farcs5 rms in each coordinate (Shaw et al. 1998).  

\section {Analysis and Results}

A global view of N157B as seen by the \Chandra\ HRCs is shown in
Fig.~1.  The low surface brightness X-ray enhancement on scales up to
$\sim 2^\prime$ is the previously resolved emission (WG98). The
spectral analysis of the \rosat\ and \asca\ data indicates that the
enhancement most likely represents the blastwave-heated thermal gas
with a characteristic temperature of $\sim 0.6$ keV (WG98).  A
spectrum from the {\sl XMM-Newton} X-ray Observatory has confirmed the
presence of the thermal component (Dennerl et al. 2001). The true
boundary of the SNR is uncertain, however. As shown in Fig. 2, the
relatively bright X-ray emission region is confined within the optical
filaments, F1, F2, and F3. But the X-ray emission clearly extends
beyond F4. These filaments, corresponding well with high velocity
features in long-slit optical spectra, are considered to be the edges
of the SNR (Chu et al. 1992). The region is also bright in radio
continuum (Fig. 3a).  The lack of enhanced X-ray emission in the
southern part of the region may be due to the presence of a foreground
dark cloud, evident in the {\sl HST} WFPC2 image (Fig. 2). This
complexity of the region is at least partly due to the coexistence of
the SNR with the OB association LH99 (Lucke \& Hodge 1970). Some of
the features may result from the photonization and/or mechanical input
from massive stars. There may be stellar wind bubbles and other 
SNRs. The interaction of the SNR N157B with this complex may also play
an important role in defining the overall X-ray morphology of the
region.

\begin{figure} 
\centerline{ {\hfil\hfil
\hfil\hfil}}
\caption{{\sl HST} WFPC2 V-band (filter F606W) image of of SNR N157B, 
overlaid by the same {\sl Chandra} X-ray intensity contours as in Fig. 1. 
Various optical features (prefixed by ``F'') as well as 
the pulsar \psr\ and the X-ray knot, as discussed in the text, 
are marked. The image shows no evidence for the optical counterpart of the 
pulsar.
\label{fig2}}
\end{figure}

\begin{figure} 
\centerline{ {\hfil\hfil
\hfil\hfil
\hfil\hfil}}
\caption{ (a) A close-up of the X-ray-emitting PWN and the associated radio 
continuum emission in SNR N157B. The X-ray contours are the same as in 
Fig. 1, but starting at the 6th level. The gray-scaled radio continuum map 
is at 3.5 cm and with a resolution of 2\farcs7 
$\times$ 1\farcs8 (HPBW) (Lazendic et al. 2000). 
(b) HRC count distribution in the same field;
various features, discussed in the text, are outlined. 
\label{fig3}}
\end{figure}

Fig. 3 presents a close-up of the high X-ray surface brightness region
of N157B along with a recent radio continuum map for comparison.  The
main X-ray emission for the region, referred to as a ``comet-shaped''
feature in WG98, is oriented SE-NW and is now fully resolved in the
\chandra\ images. As can be see in Fig. 3, embedded in the SE end
of the ``comet'' lies a bright, compact nebula (the ``head'' in WG98).
This compact nebula is elongated in a direction nearly perpendicular
to the fainter, diffuse comet tail-shaped feature. In the middle of the
bright nebula is a point-like source, representing the pulsar
\psr. The X-ray centroid of the source is at R.A., Dec (J2000) =
$5^h37^m47\fs36, -69^\circ 10^\prime$20\farcs4, consistent with the
\rosat\ position within its uncertainty. Interestingly, the
enhanced radio emission matches morphologically 
the X-ray boundary at the opposite end
of the ``comet''.  Table 1 summarizes the properties of these
individual components of N157B.

Following the procedure outlined in Gotthelf \& Wang (2000), we
conducted period folding and phase-resolved imaging analyses of the
HRC-S data. We detected a highly significant signal ($\sim 13 \sigma$)
near the expected period, at $P = 16.120868(1)$ ms for Epoch
51756.291036808 MJD.  We assumed a period derivative of $\dot P =
5.17\times 10^{-14}$ for the Epoch of the data based on a phase-connected
ephemeris derived from {\it RXTE} monitoring observations which span
the \chandra\ Epoch (Middleditch 2001, personal communications). 
Our period measurement is in excellent
agreement with the {\it RXTE} prediction, well within the expected
uncertainty, confirming the  timing accuracy of
\chandra. Fig. 4 shows the resultant light-curve folded at the peak
period. Consistent with the previous detections, the full width of the
pulse is only $\sim 10\%$.

\begin{figure} 
\centerline{ {\hfil\hfil
\psfig{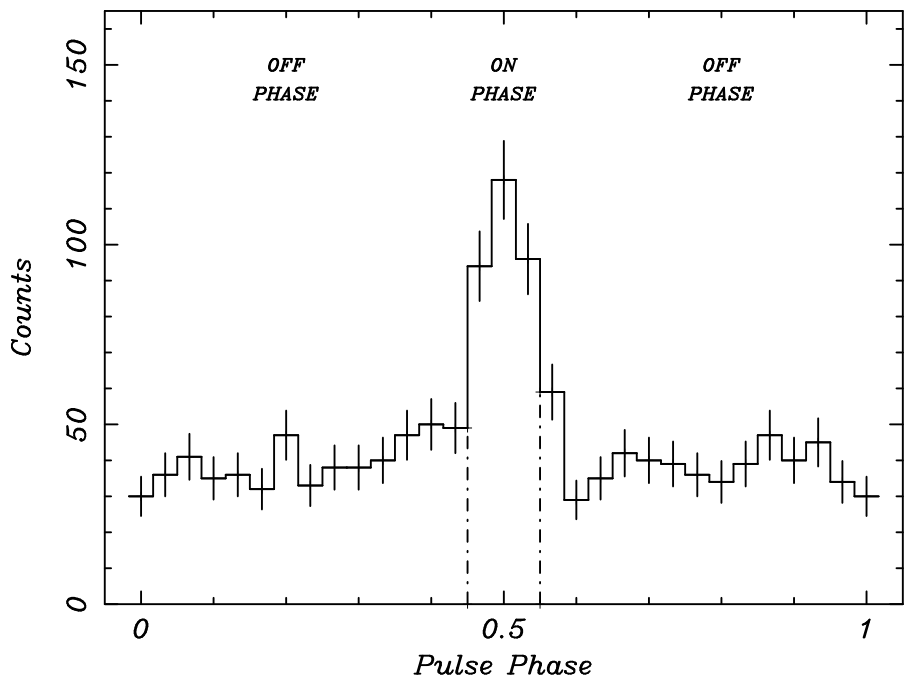}
\hfil\hfil}}
\caption{HRC-S light curve for \psr\ folded at the
ephemeris given in the text. The light curve has been extracted from a 
1\farcs3 radius aperture centered on the pulsar.
\label{fig4}}
\end{figure}

\begin{figure} 
\centerline{ {\hfil\hfil
\psfig{figure=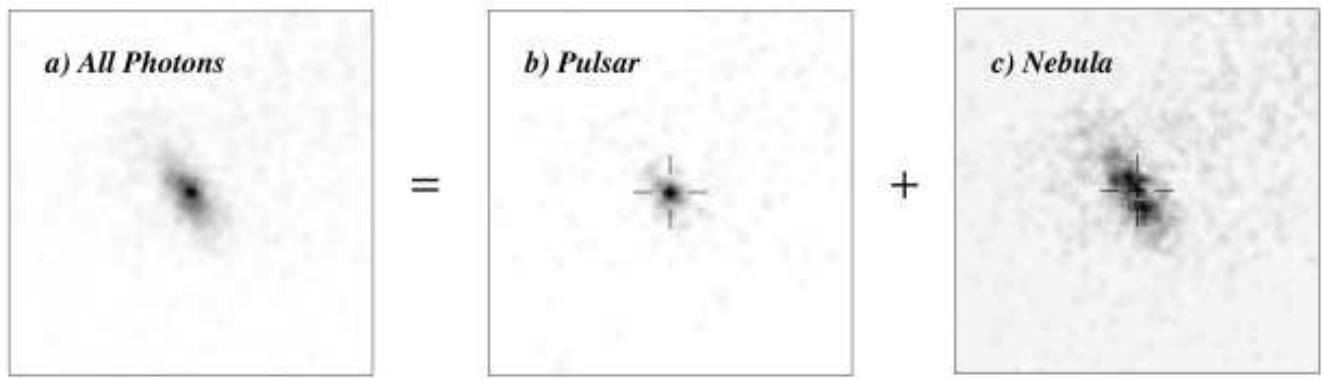,height=2.in,angle=270, clip=}
\hfil\hfil}}
\caption{Phase-subtracted X-ray intensity images of \psr:  a)
image of the pulsar and nebula (all photons) before phase-subtraction; 
b) the pulsar image containing the pulsed photons only, representative 
of the HRC point-spread function and consistent with the
ground calibration; c) same region after subtracting the
pulsar's contribution to the total flux (see text for details),
providing a good estimate of the nebula emission surrounding the
pulsar. The cross marks the pulsar's centroid. The three maps are
identically sized and linearly scaled in intensity.
\label{fig5}}
\end{figure}

The phase-resolved imaging capability of the data
allowed us to decompose the pulsed and unpulsed components. We
constructed on- and off-pulse images (Fig. 5) by selecting data in the
phase interval $0.45-0.55$ and
the rest of the pulse profile (Fig. 4). The on-pulse image 
with the phase interval-normalized off-pulse image subtracted
reproduced the expected PSF. 
The image was then scaled by a factor of 1.6 and subtracted from the 
off-pulse image to construct an image of the extended nebula emission 
(Fig.~5c) alone. 
This scaling, which minimizes the point-like contribution at 
the pulsar position, indicates that a substantial fraction 
of the pulsar's X-ray emission 
may be unpulsed (Table 1) or that considerable amounts of X-ray
emission arises in the region very close to the pulsar ($\lsim 0\farcs5$),
mimicking a point-like source. Fig. 5 illustrates a clear 
distinction between the spatially extended emission from the nebula and the 
point-like pulsed emission from the pulsar. This distinction is also 
apparent in Fig.~6, in which we present average radial intensity 
distributions around the pulsar.

\begin{figure} 
\centerline{ {\hfil\hfil
\psfig{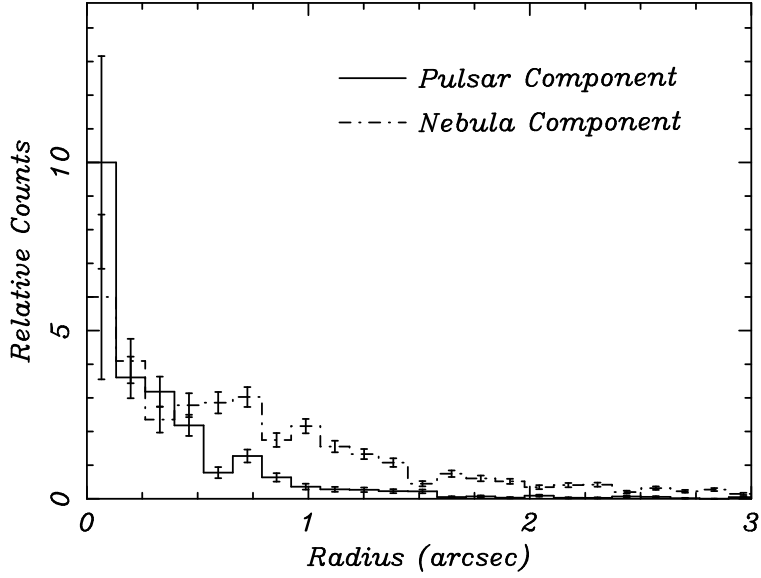}
\hfil\hfil}}
\caption{HRC-S radial intensity distributions around the point-like
emission peak of \psr. The nebula profile
(dash-dotted histogram) is significantly extended, compared with
the on-pulse profile (the solid histogram). 
\label{fig6}}
\end{figure}
\begin{figure} 
\centerline{ {\hfil\hfil
\psfig{figure=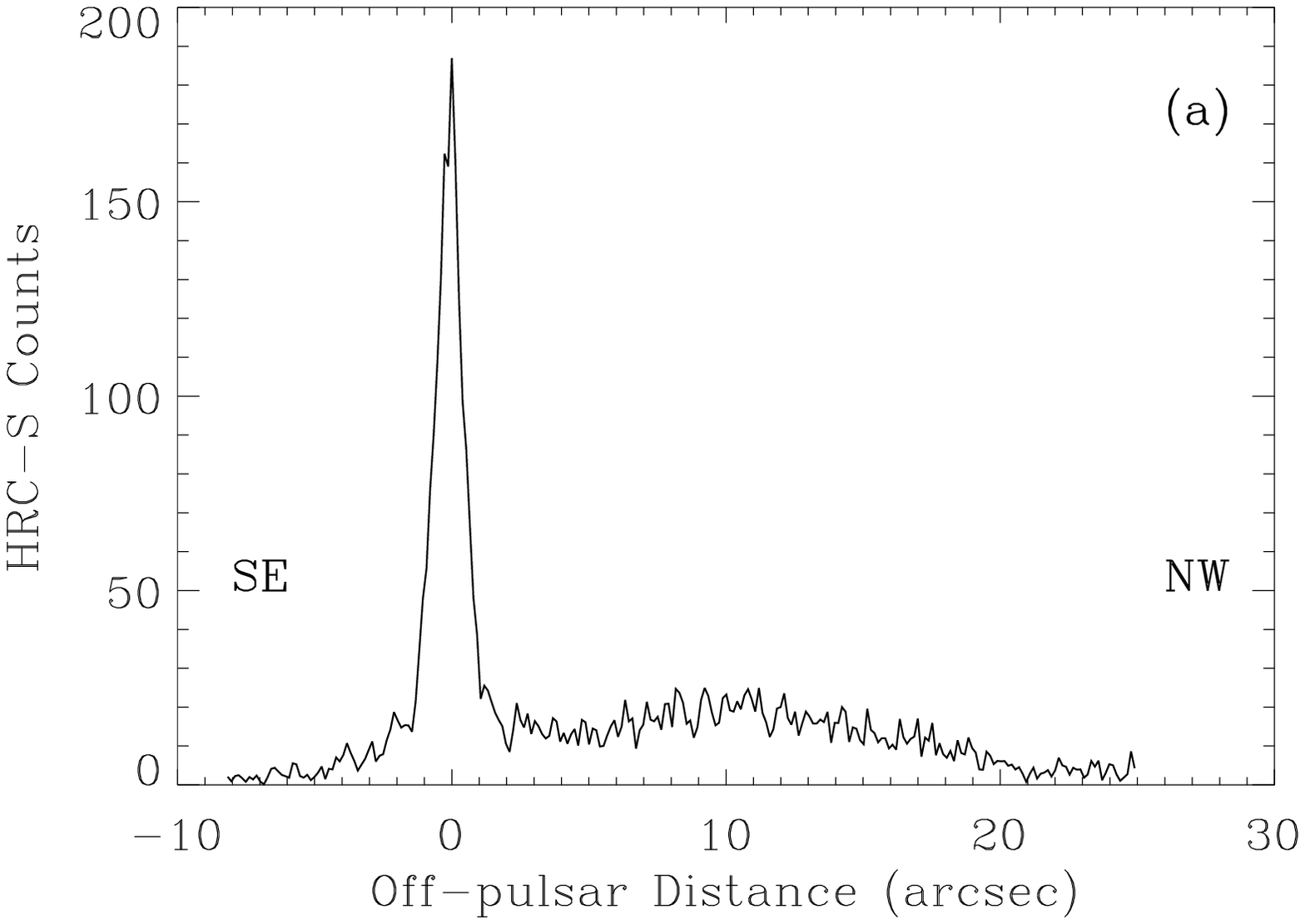,height=2.5in,angle=0, clip=}
\hfil\hfil
\psfig{figure=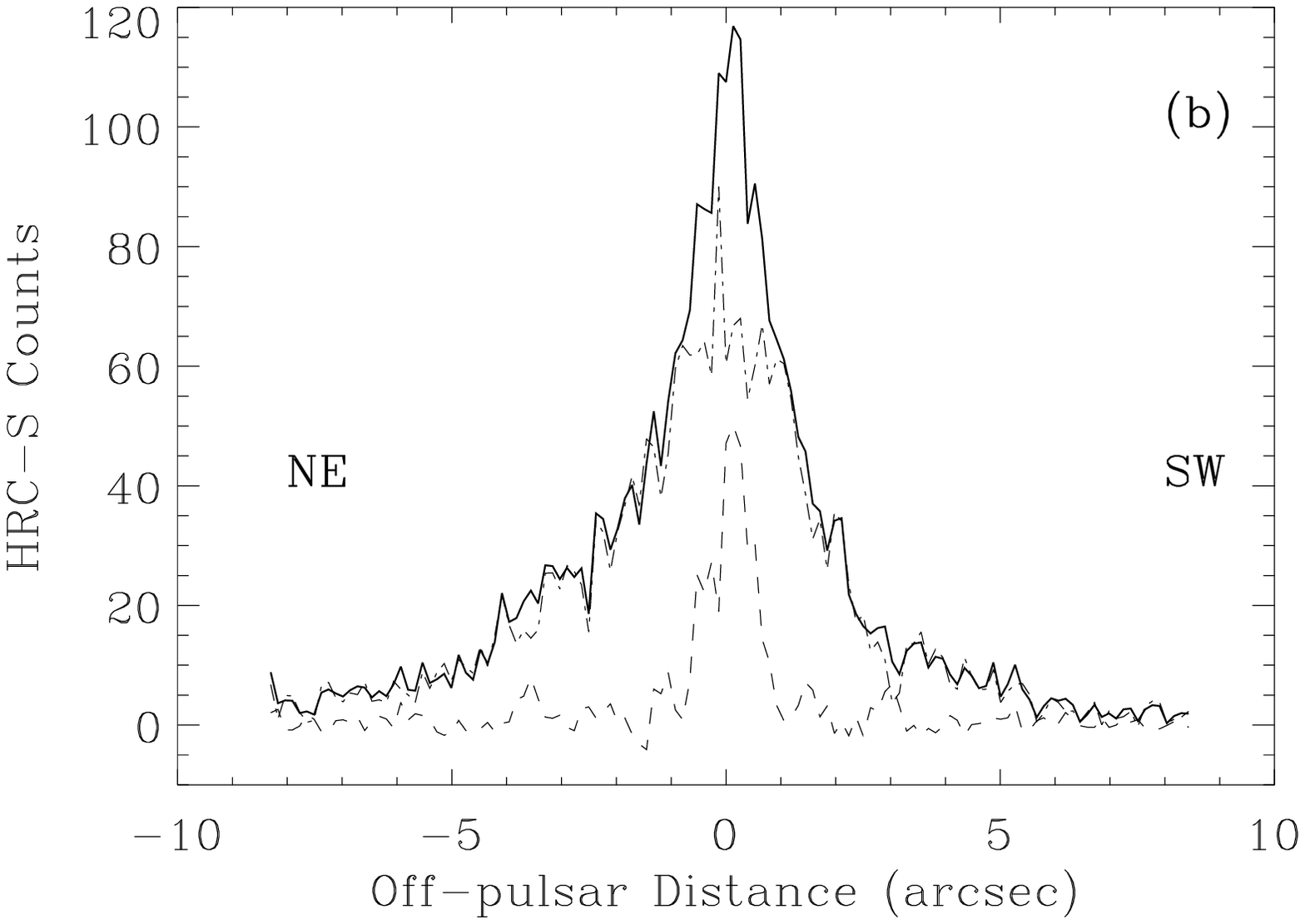,height=2.5in,angle=0, clip=}
\hfil\hfil}}
\caption{HRC-S intensity distributions along the two rectangular
cuts as outlined in Fig. 3b. The horizontal axis marks the offset
from the pulsar: (a) for the SE-NW oriented Diffuse Emission Cut
 and (b) for the SW-NE elongated Pulsar and Nebula Cut (Fig. 3b).
The solid curves represent the total intensity distributions. 
The dashed and dot-dashed curves in (b) illustrate the decomposition
of the point-like and extended nebula contributions
as discussed in the text.
\label{fig7}}
\end{figure}

Figs. 7a,b further demonstrate the average intensity distributions
along two rectangular cuts. We choose our first box (2\farcs5 $\times
15^{\prime\prime}$) to show the X-ray intensity along the long axis of
the pulsar nebula. The position angle of the box is 39.2 degs (counterclockwise
from the north; Fig. 3b). The long axis of the second box (33\farcs5
$\times 15^{\prime\prime}$) is perpendicular to the first, to show the
profile of the comet tail-shaped diffuse feature.  Although the sharpness
of the peaks is somewhat damped by the averaging, the figures allow
for both appreciating the relatively intensities along the cuts and
defining the boundaries of the three distinct components: the
point-like source, the SW-NE elongated compact nebula around the
source, and the comet tail-shaped large-scale diffuse
feature (Table 1).  In particular, the flat top plateau of the
point-source subtracted nebula intensity profile has a half width of
$\sim$ 1\farcs2 (e.g., Fig. 7b).  This width is not particularly
sensitive to the uncertainty (a couple of tens percent) in the scaling
of the above point-like source subtraction.

\begin{deluxetable}{lccc}
\tablewidth{0pt}
\tablecaption{Components of \snr
\label{tbl-1}}
\tablehead{
\colhead{Component} & \colhead{Count Rate$^a$}& $L_x^b$ &  \colhead{Size and Shape}
}
\startdata
Pulsar \\
\quad pulsed &  5.8&0.8  & point-like \\
\quad unpulsed&  9.2&1.2   & '' \\
Nebula (Torus)      &  44&8.6   & $2.5^{\prime\prime} \times 7^{\prime\prime}$ 
ellipse\\
Diffuse Feature$^c$          &  71&15    & $15^{\prime\prime} \times 30^{\prime\prime}$ ellipse \\
SNR Shell  &  150  &17& $\sim 1^{\prime}$ radius\\
\enddata
\tablenotetext{a}{in units of ${\rm~counts~ks^{-1}}$.}
\tablenotetext{b}{The luminosities are in units of $10^{35}{\rm~ergs~s^{-1}}$
and in the 0.2-4 keV band (for ease
of comparison with previous measurements; e.g., Chevalier 2000 and references
therein). The conversion from the count rate to the luminosity 
 assumes an intrinsic X-ray spectrum as a power law 
with an energy slope of 0.6 and 1.5 for the pulsar and nebula components 
(Marshall et al. 1998; WG98) and as an optically thin thermal plasma with 
a temperature of 
0.6 keV for the SNR shell. An absorbing gas column density
$10^{22} {\rm~cm^{-2}}$ (40\% solar abundances; WG98) is also assumed.}
\tablenotetext{c}{Excluding the contribution from the nebula (torus).}
\end{deluxetable}

\section {Comparison with Other Crab-like SNRs}

	A comparison of N157B with the Crab Nebula and SNR 0540-693 shows interesting
similarities and distinctions. The spindown luminosities ($\dot  E$) of 
N157B, Crab, and SNR 0540-693 pulsars are comparable:
     4.8, 4.7 and 1.5 (all in units of $10^{38} {\rm~ergs~s^{-1}}$). 
The compact X-ray nebulae around the pulsars
of these remnants are also similar, in terms of both size and 
shape (Table 1), although the elongation 
of the N157B nebula is the strongest. {\sl Chandra} images of the Crab Nebula 
show a tilted X-ray torus that is interpreted as synchrotron radiation
from the shocked toroidal pulsar wind (Weisskopf et al.
2000). This structure resembles
the X-ray-emitting rings which lie around the rotation
equator of the nearby Vela pulsar (Helfand et al. 2001), although
the latter is not as energetic ($\dot E \sim 7 \times 10^{36}
{\rm~ergs~s^{-1}}$).  
The nonthermal nebula of SNR 0540-693 exhibits a very similar overall
X-ray morphology to that of the Crab Nebula, though seen at a much larger
distance (Gotthelf \& Wang 2000). Thus, the strong 
elongation of the N157B nebula likely
represents a nearly edge-on view of the shocked toroidal pulsar wind
from \psr. The ratio of the minor to major axes of the nebula (Table
1) suggests that the toroidal wind fans out primarily within an
opening angle of $\sim 40^\circ$ about the equator of the pulsar and
sweeps a total solid angle of $\delta \Omega= 4.3$ steradian.

Just like the Crab and Vela nebulae, 
the spin axis of the N157B pulsar is perhaps aligned 
approximately with the direction of its proper motion. Both the comet-like 
morphology (Fig. 3) and the position offset of the pulsar from
the radio peak in the same direction as the orientation 
indicate that the pulsar is moving 
toward SE relative to the ambient medium (WG98). The direction is
almost perpendicular to the elongation of the nebula, thus probably 
parallel to the spin axis of the pulsar.

N157B does show several distinct characteristics. First, the X-ray to 
spindown luminosity ratio of the N157B nebula ($\sim 0.2\%$) is 
much less than that of the Crab Nebula or SNR 0540-693 (both $\sim 5\%$; Chevalier 2000 
and references therein). Even with the inclusion of the emission from the 
diffuse feature (Table 1), the ratio of N157B is still only about 0.5\%. 
 Second, the compact X-ray nebula in N157B is not associated with a 
distinct radio peak as in the Crab Nebula and in SNR 0540-693. 
Instead, the strong radio enhancement coincides spatially with the NW portion 
of the diffuse feature. Third, this very extended ($\gtrsim 5$ pc) nonthermal 
X-ray feature itself is the largest among all known Crab-like SNRs.
Its nonthermal nature was determined from a joint \rosat\ and \asca\ data 
analysis (WG98). This conclusion is consistent with a recent off-axis 
\chandra\ ACIS image 
of the region, showing that the feature is prominent at energies greater than
2 keV (L. Townsley, 2000; private communications).  
The radio emission from the region is
polarized (Lazendic et al. 2000), again indicating a nonthermal origin.
 What do these distinct characteristics tell us about N157B? 

\section{Nature of the N157B nebula}

The bright, symmetric X-ray nebula surrounding the N157B pulsar
evidently represents a PWN. In the following we argue
that this nebula is pressure-confined as proposed in WG98.  From
Fig. 3a, it is seen that there is a $\sim 20''$ 
 offset between the pulsar and the radio peak, which presumably
represents the synchrotron radiation from accumulated 
pulsar wind material. This
offset, together with the age of the pulsar and the SNR (Marshall et
al. 1998; WG98), suggests a proper motion of the pulsar as $v_p \sim
600 {\rm~km~s^{-1}}$, which alone is greater than the sound speed of
the SNR hot gas with a characteristic temperature of $\sim 0.6$ keV
(WG98). Since SNR N157B coexists with a massive star forming region, the 
medium around the pulsar may be very clumpy, as indicated in Fig. 2.
The pulsar could be moving through a local cooler and denser environment. 
In any case, it is the combination of the pulsar velocity and the 
environment that determines the morphology and X-ray emission
of the PWN. The nebulae in both Crab and SNR 0540-693 apparently lack the
right combination for them to be ram-pressure confined PWNs.

Let us now test whether this ram-pressure confined PWN interpretation
could explain the observed X-ray morphology and luminosity of the
N157B nebula.  We can define a characteristic scale for a ram-pressure
confined nebula by considering the shock radius of the pulsar wind (if confined
primarily within the solid angle $\delta\Omega$), $r_s \sim (\dot
E/\delta\Omega v_p^2 n_a \mu)^{1/2} \sim (0.3 {\rm~pc}) (v_p/600
{\rm~km~s^{-1}})^{-1}$ $ (n_a/0.6 {\rm~cm^{-3}})^{-1/2}$, where $n_a$ and
$\mu$ are the density and mean particle mass of the ambient hot gas
(WG98).  The corresponding angular size of 1\farcs2 is consistent with
the half width of the flat top intensity distribution along the major axis of 
the point-source subtracted nebula (Fig. 7b; \S 3).  
Fig. 3a shows that the X-ray intensity contours, except for the highest levels,
 are largely asymmetric, relative to the major axis of the nebula. This is
the morphological characteristics of a ram-pressure confined PWN. 
If the pulsar is moving towards SE along the minor axis, as 
is assumed in our interpretation, some asymmetry in the
intensity distribution about the major axis might also be expected even 
at the highest levels, because the reverse shock should be closer to 
the pulsar (thus stronger) on the SE side than on the NW side of the nebula. 
This of course assumes a symmetry in the pulsar wind about the major axis. 
Fig. 7a shows indication of this asymmetry: there is a separate intensity
peak, about 0\farcs5 southeast of the primary peak (pulsar). The reverse 
shock on the SE side is expected to be closer to the pulsar than in the 
direction of the major axis. Therefore, the reverse shock in the direction 
of the pulsar motion may not be fully resolved. With uncertainties in 
both the counting statistics and the PSF of the instrument, however, the data 
do not allow for a clean separation of the shocked particle emission 
from the pulsar. The relation between
the inner and outer radii of the nebula depends on the detailed MHD
dynamics of shocked pulsar wind material, which is greatly uncertain
theoretically. We thus adopt a characteristic path-length of particles
before being driven out from the nebula as $l_p \sim 0.6$ pc, the
average of the observed inner and outer radii of the nebula. The
escaping time of the particles from the nebula is then $\tau_e \sim
l_p/c_s \sim 3$ yrs with $c_s = c/\sqrt 3$, the sound velocity in the
relativistic plasma.

Following Chevalier (2000), we can estimate the synchrotron luminosity
of the confined PWN, or the bow shock region of the extended diffuse
feature. Specifically, we assume 1) the pulsar spindown power is
equally divided between magnetic fields and particles in the shocked
wind material, 2) these particles acquire a power law energy
distribution with an index $p$ and the lower energy bound is
determined by the pulsar wind speed (or the Lorentz factor
$\gamma_w$), and 3) the energy is radiated at the critical energy
$\varepsilon \propto \gamma^2 B$, where $\gamma$ is the Lorentz factor
of a particle and $B \sim 2.3 \times 10^{-4}$~G is the magnetic field
(assuming the equipartition). The integrated synchrotron spectrum is
then a power law with an energy slope of $\alpha = (p-1)/2$.  Assuming
$\alpha \sim 1.5$, the observed slope of the nonthermal emission from
N157B (WG98) indicates $p \sim 4$.  The synchrotron emission from the
ram-pressure confinement nebula is limited by the above particle
escaping time scale $\tau_e$, which is shorter than the X-ray-emitting
life-time of the particles $t_x \sim (40 {\rm~yrs}) \varepsilon^{-1/2}
(B/10^{-4}{\rm~G})^{-3/2}$, where $\varepsilon$ is in units of keV.
From the observed 0.2-4 keV band luminosity ($8.6 \times 10^{35}
{\rm~ergs~s^{-1}}$; Table 1), we infer $\gamma_w \sim 5
\times 10^6$. This estimate of $\gamma_w$ does depend on the assumed
parameters, especially $\alpha$. Based on the X-ray spectrum of an
integrated spectrum of N157B from the {\sl XMM-Newton} X-ray
Observatory, Dennerl et al. (2001) obtain a steeper power law component
with $\alpha \sim 1.8$, depending on their preferred spectral
decomposition of thermal and nonthermal components.  Adopting this
$\alpha$ value gives $\gamma_w \sim 8 \times 10^6$. Of course, the
$\alpha$ may vary with the position in the nebula, as the spectral
steepening for synchrotron burn-off is expected (e.g., Slane et
al. 2000). Assuming $\alpha \sim 1$, a typical value for Crab-like
nebulae, we find $\gamma_w \sim 2 \times 10^6$. Thus within such an
uncertainty, our inferred $\gamma_w$ is consistent with the value
($\sim 3 \times 10^6$) estimated from the detailed modeling of the
Crab Nebula (Kennel \& Coroniti 1984).

In contrast, if the nebula in N157B were a normal PWN as in the Crab
Nebula or SNR 0540-693, the inferred $\gamma_w$ would be much smaller.
By comparing the observed luminosity with the integration of Equation
9 of Chevalier (2000) over the $0.2-4$ keV range, we find $\gamma_w
\sim 1 \times 10^5$, which is more than one order of magnitude lower
than the inferred for the Crab pulsar. In this case, the synchrotron 
radiation of the particles is {\sl not} limited
by the escaping time as is in a ram-pressure confined PWN, but the observed 
X-ray luminosity represents only the high energy tail of the radiation 
spectrum. Most importantly, this PWN
model of the N157B nebula would offer no explanation for the trailing
 {\sl nonthermal} diffuse X-ray feature.

The ram-pressure confined PWN scenario provides a natural explanation
for the nonthermal diffuse X-ray feature trailing the
pulsar as well as its associated radio emission. The nonthermal
spectrum represents the continuing synchrotron cooling of the
particles escaped from the nebula. The pressure outside of the compact
nebula should be comparable to that of hot gas in the SNR. Further
assuming equipartition, we estimate the magnetic field as $\sim 3
\times 10^{-5} $G, about an order of magnitude lower than in the
nebula. Correspondingly, the X-ray-emitting lifetime of the particles
is a factor of $\sim 21$ longer than for those in the compact
nebula. The length of the diffuse emission feature ($\sim 10$ pc)
requires a particle transporting velocity of $\gsim 10^4
{\rm~km~s^{-1}}$. This velocity can be achieved easily by the bulk
outflow of shocked pulsar wind material from the compact nebula, but
would be difficult for other mechanisms (e.g, diffusion).
Furthermore, the bulk outflow is naturally one-sided, as observed.
The outflow, after being terminated, feeds a bubble of pulsar wind
material, which may be responsible for part of the diffuse emission
feature (WG98). The observed radio peak represents the emission from
pulsar wind material accumulated during the lifetime of \psr, whose
age is smaller than the lifetime of radio-emitting particles. The low
efficiency of radio emission, compared with X-ray radiation, naturally
explains the lack of a strong radio peak at the nebula close to the
pulsar, which has probably moved outside the bubble. The nebula 
represents the emission only from fresh particles
just out of the pulsar.

	In short, the ram-pressure confined PWN model
is consistent with the existing data on N157B and provides a working
scenario for understanding the pulsar wind and its interaction with
the ambient medium. A detailed spatially-resolved X-ray spectroscopy,
which can be provided by a deep on-axis \chandra\ ACIS observation,
will be most useful for further testing the model, for tightening
constraints on various parameters of the pulsar wind (e.g.,
$\gamma_w$) and the shock acceleration of particles (e.g., $p$), and
for determining the nature of various diffuse X-ray emission features of
the SNR.

\acknowledgements

We thank R. Chevalier and the anonymous referee for their comments and 
gratefully acknowledge the {\sl Chandra} team, especially Steve Murray,
for arranging the re-observation of N157B and for substantial help during
the course of this work, which was funded by SAO {\sl Chandra} grant 
GO0--1124 and NASA LTSA grant NAG5--7935.
\vfil
\eject

\end{document}